# INHERENT LIMITATIONS OF ICE CUBE WITH REGARDS TO BOTH CP VIOLATION PHYSICS, AND DATA COLLECTION

## A.W. Beckwith

*E-mail: abeckwith@UH.edu*

**Abstract.** We examine how ICE CUBE is impacted via definite observational limits, a potentially serious flaw in estimated collection ability as predicted by a short fall in Monte Carlo simulations, and a datum that the CP violations connected with Baryogenesis are not predicted by SUSY accurately. All these considerations, plus the known datum that water is a superior medium for data collection than block ice argue pervasively in first building neutrino physics collection systems first in the Mediterranean sea , and relying upon ICE CUBE as a back up to planned and running systems in place near Europe. This is made urgent also in the known datum that both the near European sub continent facilities now have bio luminence problems in the water, but have the same sensitivity as the South polar region neutrino astrophysics facilities now planned for

# INTRODUCTION

In this paper, we will discuss several layers of doubt as to the technical feasibility of the ICE CUBE experimental system. This is in part due to a presentation by the author at the 2nd International Summer School on Astro particle Physics at NIJMEGEN. None of this is to suggest that a system purporting to do the job ICE CUBE is attempting to do, i.e. dramatically improved neutrino physics, as well as investigations of the state of the early universe with respect to phase transitions, baryogenesis, and CP violations should be abandoned. We should do these things. However, as presented below, there are serious problems with the present system as planned which need to be re evaluated beyond the normal imperative of funding agencies which need project work in order to justify their existence. Too much is at stake here. And what is being presented is a cautionary tale of what needs to be re evaluated before we commit ourselves to an optimistically stated update of the Armanda system in the South Polar regions.

## OBSERVATIONAL PROBLEMS INHERENT IN GEOGRAPHICAL LOCATION.

For one thing, the South Polar Region is, with respect to neutrinos, is largely incapable of observing the Milky Ways profile from a neutrino data stand point. This is the opposite of standard visual radiation, which in fact favors the South Polar Region from an optical radiation stand point over the North Pole. This odd datum was commented upon in the 2nd International Summer School on Astro particle Physics at NIJMEGEN, as well as a presentation by Dr. Carr[1] who listed the advantages of non Antarctic based neutrino collection systems as follows:

1) Water based collection systems purportedly have superior angular resolution properties over ice based collection systems. This is due in part to ice creating far more light scattering than water. Ice has dust layers on its surface, and in addition, in its interior has bubbles; while on average water is far more homogeneous in permeability properties.

2) ICE CUBE is designed to have the same sensitivity as Km3NET as far as detection of gamma ray bursts, and other neutrino generating events.

3) The main shortfall of the PRESENT water based neutrino systems evidently is in shallow to medium dept bio luminescence of the water. A problem which Km3NET is designed to circumvent due to being an extremely deep water system.

4) In addition, Dr. Carr, and others noted that ICE CUBE cannot get general neutrino physics data from the Milky Way galactic center, while water based systems far closer to the equator of Earth have no such limitations[2].

If the above is true, it stands to reason that the ANTARES replacement, Km3NET, should be put in first, and a significant re think of what the scientific community wants from ICE CUBE should commence. However, this is not the only problem. See below. We will next present evidence to the effect that even if the above considerations mentioned already were dealt with, that the Monte Carlo simulation used to justify ICE CUBE may be invalid.

## MONTE CARLO SIMULATION SHORTFALL.

(1) In astro physics 20 (2004), pp 507 - 532, there is a diagram of ice cube, in figure 1, page 510

(2) 1st problem; the simulated Monte Carlo problem for ice cube is not same as the planned installation .The only thing the same is spatial separation of strings placed 125 meters apart from each other.

(3) 2nd problem ; A method of calculating sensitivity given by Gary Feldman, and R. Cousins in Phys Rev D. vol 57, No. 7, APRIL 1ST, 1998, produced a method to quantify "sensitivity" of an experiment via an upper limit, $\bar{\mu}$ , calculated in the absence of a signal.

(3a) We get $\bar{\mu}_{90}$ from the mean number of background events over all limits obtained from all experimental outcomes.

(3b) Feldman and Cousins in their article used $\langle n_s \rangle$ as at least = $\bar{\mu}_{90}$ = # of signal events

(3c) From 90% c.l. average factor, they defined a "model rejection factor" (MRF) for an arbitrary source source spectrum $\phi_S$

Therefore, MRF = $\bar{\mu}_{90} / \langle n_s \rangle$

Claim: [$\phi_{90}$ = # of expected events] = [$\phi_S$ times MRF = average upper limit]

Question: If Antarctica is as cold as it is 1/2 a year, is it possible to get a valid "mrf" factor which permits a suitable number of events simulated in a realistic manner? It is very doubtful.

(3d) Ice cube is supposed to work for 15 years to get sufficient data.

Question: Is this feasible if the real- de facto "rmf" due to bad weather, etc, makes the real detection we can realistically expect 1/10 of what Monte Carlo simulates?

Consider if one will, that if the data collection time length is inaccurate, and at least twice as long, that the entire ICECUBE system may be rendered redundant with respect to newer systems which may have dramatically improved technical systems.

Again though, assuming that even if the Monte Carlo detection rate is not wrong, there is yet another issue. One which raises the question of if or not ICECUBE is trying to falsify a paradigm which is, at least from the particle physics stand point on shaky ground.

## BARYOGENESIS PROBLEMS, AND THE ABSENCE OF ANTI MATTER

(1) Ice Cube purports to be investigating SUSY, and the existence of topopological defects experimentally created

In the Feynman festival III, C Rangacharyulu told me that a SUSY derived from the standard model was really inappropriate for explaining baryogenesis, i.e. the dearth of anti matter in the universe.

Theoretical: $\dfrac{\eta_\beta - \eta_{\bar{\beta}}}{\eta_\gamma} \cong 10^{-18}$

Experimental: $\dfrac{\eta_\beta - \eta_{\bar{\beta}}}{\eta_\gamma} \cong 10^{-10}$

This implies CP violations beyond the standard model. If the SUSY being planned for falsifiable measurement procedures with respect to ICE Cube is similar to the SUSY which is shown as inadequate for Baryogenesis, then ICE Cube is trying to set up falsifiable predictions for paradigm already discredited. Phys Rev D 73, 072005(2006), pp 1-43. "Search for T violating transverse muon production in the $K^+ \rightarrow \Pi^0 \mu^+ \nu$ decay"

# CONCLUSION

The evidence so presented is not a refutation of the basic mission of ICE CUBE. However, before spending the money, we really need to prioritize what we are doing in lieu of some known datum

1) Water based collection systems are supposed to work in tandem with ICE CUBE. Armanda, in particular needs to be replaced due to bio luminescence problems in water which are degrading the effectiveness of the system, whereas the planned sensitivity of Km3Net , in deep water in the Mediterranean is at least on the same level as ICE CUBE, even with an optimistic reading as to the efficiency of ICE CUBES detectors.

2) There are sound statistical reasons to doubt ICE CUBE will be able to get adequate data for its evaluative mission in 15 years. This is based upon short falls in the Monte Carlo system discussed above.

3) The known datum that particle physicists are about ready , in 2007 , to use advanced Kaon physics experiments, as outlined in a Physical Review D article published in 2006 to fully investigate CP violations and to investigate physics well beyond the SUSY models we presently work with (which assign a duality between fermions and bosons equivalent to particles and anti particles. i.e. quarks ⟵⟶ squarks, etc ) in Japan for several years brings into question if or not SUSY, and CP violations as visualized by the designers of ICE CUBE are the right models to be investigated for falsifiable experimental predictions[3] . The SUSY predictions are to be used to investigate effects beyond the electro weak forces of nature and possible phase transitions in the early universe. We may well be better off if we concentrate upon building a functional Km3Net test bed and getting it working, and also attempting to understand the results of the fore coming extensions of the 'Muon polarization in the $K^+ \rightarrow \Pi^0 \mu^+ \nu$ decay" experiments to understand if or not we can measure CP violation and early universe phase transitions via ICE CUBE.

4) Finally is the datum of extreme weather. It is not to be underestimated. In Siberia, Russians have to take batteries out of cars in the dead of winter and take them inside their houses in order to keep them from being ruined. Certainly ARMANDA dealt with cold weather issues fairly well with robust technology. ICE CUBE though is projecting a 15 year run, MINIMUM, in order to obtain a suitable assembly of 'good' data for evaluation. This is an order of magnitude beyond the ARMANDA experiments duration.

# BIBLIOGRAPHY

---

[1] http://nijmegen06.astro.ru.nl/docs/Carr-1.ppt#686,3,Multi-Messenger Astronomy; http://nijmegen06.astro.ru.nl/docs/Carr-2.ppt

[2] Comments from Dr. Carr and others about Neutrino telescopes at Nijmegen06 in response to questions from the audience

[3] Comments from C Rangacharyulu at the Feynman III festival in response from questions from the audience : http://www.physics.umd.edu/robot/ff3